\documentclass[prd,eqsecnum,showpacs,showkeys,floatfix,nofootinbib,preprint,
               fleqn,tightenlines]{revtex4}

\usepackage{amsmath}
\usepackage{amsfonts}
\usepackage{latexsym}
\usepackage{graphicx}
\usepackage{mathrsfs}

\newcommand{\version}{v5}

\newcommand{\mathindentnum}{1cm}               

\newcommand{\alphatwoflavor}{\gamma}           
\newcommand{\Deltabzerotwoflavor}{\Delta b}
\newcommand{\Deltamsquaretwoflavor}{\Delta m^2}
\newcommand{\widebar}{\overline}               
\newcommand{\beq}{\begin{equation}}
\newcommand{\eeq}{\end{equation}}
\newcommand{\beqa}{\begin{eqnarray}}
\newcommand{\eeqa}{\end{eqnarray}}
\newcommand{\no}{\nonumber}

\def\e{{\rm e}}

\newcommand{\half}{{\textstyle \frac{1}{2}}}

\renewcommand{\i}{{\rm i}}

\newcommand{\SM}   {Standard Model}

\begin{document}
\noindent Phys. Rev. D 71, 113008 (2005)
\hspace*{\fill} hep-ph/0504274, KA--TP--05--2005 (\version)

\vspace*{2\baselineskip}
\title{Possible energy dependence of
      $\Theta_{13}$ in neutrino oscillations}

\author{Frans R.\ Klinkhamer}
\email{frans.klinkhamer@physik.uni-karlsruhe.de}
\affiliation{Institute for Theoretical Physics,
University of Karlsruhe (TH), 76128 Karlsruhe, Germany\\}

\begin{abstract}
\noindent
A simple three-flavor neutrino-oscillation model is discussed which has
both nonzero mass differences and timelike Fermi-point splittings,
together with a combined bi-maximal and trimaximal mixing pattern. One
possible consequence would be new  effects in $\nu_\mu \to \nu_e$ oscillations,
characterized by an energy-dependent effective mixing angle $\Theta_{13}$.
Future experiments such as T2K and NO$\nu$A,
and perhaps even the current MINOS experiment,
could look for these effects.
\end{abstract}


\maketitle

\section{Introduction}
\label{sec:intro}

The standard explanation of neutrino oscillations relies on mass
differences \cite{GribovPontecorvo,BilenkyPontecorvo1976,BilenkyPontecorvo1978} and
two recent reviews can be found in Refs.~\cite{Barger-etal,McKeownVogel}.
There have, of course, been many suggestions for alternative mechanisms; see, e.g.,
Refs.~\cite{Gasperini,Halprin,ColemanGlashow,Huber-etal,KosteleckyMewes,BarenboimMavromatos}
and references therein.
One possibility motivated by condensed-matter physics would be
Fermi-point splitting of the \SM~fermions
due to a new type of quantum phase transition
\cite{KlinkhamerVolovikJETPL,KlinkhamerVolovikIJMPA}.

The phenomenology of a ``radical'' neutrino-oscillation model with strictly
zero mass differences and nonzero Fermi-point splittings has been discussed in
Refs.~\cite{KlinkhamerJETPL,Klinkhamer0407200}.
A less extreme possibility would be a type of  ``stealth scenario''
with Fermi-point-splitting effects hiding behind mass-difference
effects \cite{Klinkhamer0407200}. The present
article aims to give an exploratory  discussion of this last possibility.

Specifically, we consider the case with both mass-square differences
$\Delta m^2$ and \emph{timelike} Fermi-point splittings $\Delta b_0$.
(Timelike Fermi-point splittings preserve spatial isotropy, whereas
spacelike splittings give anisotropic neutrino
oscillations \cite{KlinkhamerJETPL}.)
There are then two methods to determine the presence of these $\Delta b_0$ terms.
The first method is to use sufficiently high neutrino energies $E_\nu$ so that
the mass-difference effects drop out,
$|\Delta m^2/(2E_\nu)| \ll |\Delta b_0|$ for generic mixing angles.
The second method is to look at a particular  process which is expected to
be small for standard mass-difference neutrino oscillations.

An example of the second method is provided by the
standard appearance probability
$P(\nu_\mu\rightarrow \nu_e) \propto \sin^2 (2\theta_{13})$,
for small values of the mixing angle $\theta_{13}$ and not too large
travel distance $L$. If there are relatively weak
Fermi-point splittings in addition to the mass differences,
one has a similar expression  for $P(\nu_\mu\rightarrow \nu_e)$ but with
the ``bare'' mixing angle
$\theta_{13}$  replaced by
an effective mixing angle $\Theta_{13}$, which is now a function of
the basic parameters. Phenomenologically,
the most interesting consequence would be  that $\Theta_{13}$ becomes
energy dependent. The resulting
$\nu_e$ energy spectrum near the first oscillation maximum will be discussed
in detail for a simple three-flavor model. Let us remark that,
quite generally, the goal of the present article is  to point
out a possible energy dependence of \emph{all} neutrino-oscillation
parameters and the simple three-flavor model is used as an explicit example.

The outline of this article is as follows.
Section \ref{sec:two-flavor-model} gives the basic mechanism
for the case of two neutrino flavors.
Section \ref{sec:three-flavor-model} incorporates these results
in a simple three-flavor model with a single dimensionless parameter
to describe the relative strength of Fermi-point-splitting
and mass-difference effects.
Section \ref{sec:phenomenology} addresses certain phenomenological issues
and gives possible signatures for planned or proposed superbeam experiments
(e.g., T2K or NO$\nu$A).
In addition, further model results are given which may be relevant
to the current MINOS experiment.
Section \ref{sec:conclusion} contains concluding remarks.

\section{Two-flavor model}
\label{sec:two-flavor-model}
\mathindent=\mathindentnum

The starting point of our discussion is the following generalized neutrino
dispersion relation \cite{KlinkhamerVolovikIJMPA}:
\beq
E_\nu(\vec p\,) = \sqrt{(c_\nu\,p  + b_{0\nu})^2 + m_\nu^2\,c_\nu^4}
\sim c_\nu\,p  + b_{0\nu} + m_\nu^2\,c_\nu^3/(2\,p)  \,,
\label{dispersionrelation}
\eeq
for large enough neutrino momentum $p \equiv |\vec p\,|$ and with
a maximum neutrino velocity $c_\nu$,
a timelike Fermi-point-splitting parameter $b_{0\nu}$,
and a mass parameter $m_\nu$.
Now, consider two-flavor vacuum neutrino oscillations from both mass
differences and timelike Fermi-point splittings
\cite{KlinkhamerJETPL,Klinkhamer0407200}. The relevant energy
differences are then obtained from the eigenvalues of
the $2\times 2$ Hermitian matrix
\cite{ColemanGlashow}
\beq
c_\nu\,p\,A  + B_0 + M^2\,c_\nu^3/(2\,p)  \,,
\label{2flavormatrix}
\eeq
with $p \equiv |\vec p\,|$ and
$2\times 2$ Hermitian matrices $A$,  $B_0$, and $M^2$.
In our case, we have $A$
equal to the identity matrix, $A = \openone$, and the maximum neutrino
velocity $c_\nu$ equal to the velocity of light \emph{in vacuo}, $c_\nu=c$.
Natural units with $\hbar$ $=$ $c$ $=$ $1$ will be used in the following.

For two neutrino flavors (denoted $f$ and $g$) and large enough beam
energy $E_\nu$, the survival and
appearance probabilities over a travel distance $L$
are given by
\begin{subequations}\label{Ps}
\beqa
P(\nu_f\rightarrow \nu_f)  &=& 1- P(\nu_f\rightarrow \nu_g)
\,,\label{Psurv}\\[4mm]
P(\nu_f\rightarrow \nu_g)  &\sim&
\sin^2 \left( 2\,\Theta  \right) \; \sin^2 \left( \Delta E\,L/2 \right)
\,,\label{Pappear}
\eeqa
\end{subequations}
with $\Delta E \geq 0$ and $\Theta \in [\,0,\pi/2\,]$ defined by \cite{ColemanGlashow}
\begin{subequations}\label{Delta-Theta}
\beqa
\Delta E \,\sin 2\,\Theta  &=&
|\,\Deltamsquaretwoflavor/(2E_\nu) \sin 2\theta +
\exp(\i\, 2\, \alphatwoflavor)\;\Deltabzerotwoflavor\, \sin 2\chi \,| \,,
\label{Delta}\\[4mm]
\Delta E \,\cos 2\,\Theta  &=&
\Deltamsquaretwoflavor/(2E_\nu) \cos 2\theta + \Deltabzerotwoflavor\, \cos 2\chi \,.
\label{Theta}
\eeqa
\end{subequations}
Here, $\Deltamsquaretwoflavor \equiv m_2^2-m_1^2$
is the difference of the eigenvalues of the matrix $M^2$
and $\theta$ is the related mixing angle.
In the same way, $\Deltabzerotwoflavor \equiv b_0^{(2)}- b_0^{(1)}$ and
$\chi$ come from the matrix $B_0$. In addition, there is one relative phase,
$\alphatwoflavor$.

Equations (\ref{Delta-Theta}ab) make clear that,
provided $\Delta m^2$ and $\Deltabzerotwoflavor$ are nonzero, the effective
mixing angle $\Theta$ interpolates between
a value close to $\chi$ for relatively high beam energy $E_\nu$
and a value close to $\theta$ for relatively low beam energy $E_\nu$
[but still large enough for Eq.~(\ref{2flavormatrix}) to apply, see
also Sec.~\ref{sec:preliminaries}]. In order to be more explicit,
we restrict ourselves to the two ``extreme'' values for the
mixing angles $\theta$ and $\chi$, namely, the values $0$ and $\pi/4$.
We, again, assume that both $\Deltamsquaretwoflavor$ and $\Deltabzerotwoflavor$ are
nonzero. Three of the four possible cases then lead to neutrino oscillations:
\beq
\begin{array}{lcl}
\mathrm{case}\hspace{1ex} 1:\quad \theta=0     &\quad\mathrm{and}\quad&  \chi=\pi/4\,,\\
\mathrm{case}\hspace{1ex} 2:\quad \theta=\pi/4 &\quad\mathrm{and}\quad&  \chi=\pi/4\,,\\
\mathrm{case}\hspace{1ex} 3:\quad \theta=\pi/4 &\quad\mathrm{and}\quad & \chi=0\,.
\end{array}
\label{case123}
\eeq
Since the results for case 3 follow simply from those for case 1,
we only need to give the results for cases 1 and 2
(denoted by a single and a double prime, respectively).

Case 1 has appearance  probability
\begin{subequations}
\label{PandTheta-case1}
\beqa
P^{\prime}(\nu_f\rightarrow \nu_g)         &\sim&
\sin^2 \left( 2\,\Theta^{\prime}  \right) \;
\sin^2 \left(\, \Delta E^\prime\,L/2\, \right)\,,
\label{P-case1}\\[4mm]
\Delta E^\prime  &=&
\sqrt{(\Deltabzerotwoflavor)^2+(\Deltamsquaretwoflavor/(2E_\nu))^2} \,,
\label{DeltaE-case1}\\[4mm]
 2\,\Theta^{\prime}   &=& \arctan
\left(\,
\frac{|\Deltabzerotwoflavor|}{\Deltamsquaretwoflavor/(2E_\nu)}
\,\right) \,,
\label{Theta-case1}
\eeqa
\end{subequations}
with the dependence on phase $\alphatwoflavor$ dropping out
altogether. Observe that the effective
mixing angle $\Theta^{\prime}$ is energy dependent, rising from a
value $0$ at $E_\nu \sim 0$ to a value $\pi/4$ for $E_\nu \to \infty$.

Case 2 has survival probability
\beqa
P^{\prime\prime}(\nu_f\rightarrow \nu_f)         &\sim&
1 - \sin^2 \left( \,
\left|\, \exp(\i\, 2\, \alphatwoflavor) \; \Deltabzerotwoflavor+
         \Deltamsquaretwoflavor/(2E_\nu)\,\right|\,L/2\, \right)\,,
\label{P-case2}
\eeqa
with a constant mixing angle $\Theta^{\prime\prime}  = \pi/4$.
Observe that the probabilities are $\alphatwoflavor$ dependent.
In fact, neutrino oscillations  at energies
$E_\nu = |\Deltamsquaretwoflavor/(2 \Deltabzerotwoflavor)|$
are entirely suppressed for $\alphatwoflavor=\pi/2$ or $0$,
depending on the relative sign
of $\Deltabzerotwoflavor$ and $\Deltamsquaretwoflavor$.

Case 3 has a survival probability $P^{\prime\prime\prime}$ and
effective mixing angle $\Theta^{\prime\prime\prime}$
given by (\ref{PandTheta-case1}abc)
with $\Deltabzerotwoflavor$ and $\Deltamsquaretwoflavor/(2E_\nu)$ interchanged.
This implies that $\Theta^{\prime\prime\prime}$ has a
value $\pi/4$ at $E_\nu \sim 0$  and vanishes for $E_\nu \to \infty$.

With the heuristics of the two-flavor case established, we turn to the
more complicated three-flavor case in the next section.

\section{Three-flavor model}
\label{sec:three-flavor-model}

For three neutrino flavors, the diagonalization of the energy matrix
(\ref{2flavormatrix}), now in terms of
$3\times 3$ Hermitian matrices $A=\openone$, $B_0$, and $M^2$,
gives  many more mixing angles and phases than for the two-flavor case.
Using Iwasawa decompositions of the relevant $SU(3)$ matrices, one has the
following parameters:
$\{ \theta_{21},\theta_{32},\theta_{13},\delta \}$ for an $SU(3)$ matrix $X$
from $M^2$,
$\{ \chi_{21},\chi_{32},\chi_{13},\epsilon \}$ for an $SU(3)$ matrix $Y$
from $B_0$,
and $\{ \alpha,\beta \}$ for a diagonal phase-factor matrix
$P$ generated by the Cartan subalgebra.

Specifically, the matrices $P$ and $N=X,Y$ are defined as follows:
\begin{subequations}\label{PXYdef}
\beqa
P &\equiv&
\mathrm{diag}\left(\,
\e^{\i\beta}  \, , \, \e^{-\i(\alpha+\beta )}  \, , \, \e^{\i\alpha} \,
\right)
\equiv
\left( \begin{array}{ccc}
\e^{\i\beta} & 0 & 0 \\
0 &\e^{-\i(\alpha+\beta )} & 0 \\
0 &0 & \e^{\i\alpha} \end{array} \right)\,,\\[4mm]
N &\equiv&
\left( \begin{array}{ccc}
1 & 0 & 0 \\ 0 & c_{32} & s_{32} \\ 0 & -s_{32} & c_{32}
\end{array} \right)
\cdot
\left( \begin{array}{ccc}
c_{13} & 0 & \;+ s_{13}\,\e^{+i\omega} \\ 0 & 1 & 0 \\
-s_{13}\,\e^{-i\omega}& 0 & c_{13}
\end{array} \right)
\cdot
\left( \begin{array}{ccc}
c_{21} & s_{21} & 0 \\ -s_{21}& c_{21} & 0 \\ 0 & 0 & 1
\end{array} \right)\,,
\eeqa
\end{subequations}
with $\{s_{ij},c_{ij},\omega \}$ equal to
$\{\sin\theta_{ij},\cos\theta_{ij},\delta\}$ for the matrix $N=X$
and to $\{\sin\chi_{ij},\cos\chi_{ij},\epsilon\}$ for the matrix $N=Y$.
The relevant terms of the Hamiltonian in the flavor basis are then
\beq
D_p +  X\, D_{m} \, X^{-1} + P\, Y\, D_{b_0} \, Y^{-1}\,P^{-1} \,,
\label{3flavormatrix}
\eeq
with diagonal matrices
\mathindent=0cm
\beq
 D_p \equiv  \mathrm{diag}\left(\,
p  \, , \,
p  \, , \,
p
\,\right) \, , \quad
  D_{m}   \equiv   \mathrm{diag} \left(\,
\frac{m^2_{1}}{2 p} \, , \,
\frac{m^2_{2}}{2 p} \, , \,
\frac{m^2_{3}}{2 p}
\,\right) \, , \quad
 D_{b_0} \equiv  \mathrm{diag}\left(\,
b_0^{(1)}  \, , \,
b_0^{(2)}  \, , \,
b_0^{(3)}
\,\right) \, ,
\eeq
\mathindent=\mathindentnum
for $p \equiv |\vec p\,|$ and $c_\nu=c=1$.
For later use, we already define the differences
$\Delta m_{ij}^2 \equiv m_i^2-m_j^2$
and  $\Delta b_0^{(ij)} \equiv b_0^{(i)}- b_0^{(j)}$.

The structure of the vacuum  neutrino-oscillation probabilities is
the same as for the
standard case ($D_{b_0}=0$); see, e.g., Eqs.~(7)--(10) of Ref.~\cite{Barger-etal}.
These probabilities are given in terms of six parameters
$\Delta E_{21}$, $\Delta E_{31}$, $\Theta_{21}$, $\Theta_{32}$, $\Theta_{13}$,
and $\Delta$,
which are now complicated functions of the fourteen original parameters
$\Delta m^2_{21}/(2p)$, $\Delta m^2_{31}/(2p)$,
$\Delta b_0^{(21)}$, $\Delta b_0^{(31)}$, $\theta_{21}$,
$\theta_{32}$, $\theta_{13}$, $\delta,
\chi_{21}$, $\chi_{32}$, $\chi_{13}$, $\epsilon$, $\alpha$, and $\beta$.

In order to illustrate the basic idea and to use the two-flavor
results of the previous section, we consider a specific model with
the following mass-term differences and bi-maximal mixing angles:
\begin{subequations}
\label{standardmassangles}
\beqa
\Delta m^2_{21}/(2p) &=& 0\,,\quad
\Delta m^2_{31}/(2p) \equiv \Delta\mu >0 \,,\\[4mm]
\theta_{13}&=&0\,,\quad\theta_{21}=\theta_{32}= \pi/4\,,
\eeqa
\end{subequations}
a similar pattern of timelike Fermi-point splittings and trimaximal mixing angles:
\begin{subequations}
\label{FPSmassangles}
\beqa
\Delta b_0^{(21)}  &=&0\,, \quad \Delta b_0^{(31)} \equiv \Delta b_0  >0\,,\\[4mm]
\chi_{13}&=&\chi_{21} =\chi_{32}=\pi/4\,,
\eeqa
\end{subequations}
and vanishing phases:
\beq
\delta = \epsilon = \alpha=\beta =0\,.
\label{phases}
\eeq
The values (\ref{standardmassangles}ab)
are more or less standard
with $|\Delta m^2_{31}| \approx 2.5\times 10^{-3}\;{\rm eV}^2$;
cf.  Refs.~\cite{Barger-etal,McKeownVogel}.
The actual values of $\theta_{21}$ and  $\chi_{21}$ are,
in fact, irrelevant for the case $\Delta m^2_{21}=\Delta b_0^{(21)}=0$.
As it stands, the simple model depends on only one positive dimensionless
parameter, $\Delta b_0/\Delta\mu$.

Explicit expressions for the energy  eigenvalues $E_n$ of the corresponding
matrix (\ref{3flavormatrix}) are readily obtained and give the following
differences:
\begin{subequations}
\label{DeltaEmodel}
\beqa
\Delta E_{21} &\equiv& E_2 -E_1=
\half \left(\Delta\mu + \Delta b_0 -\sqrt{\Delta\mu^2 + \Delta b_0^2} \right) \,,
\label{DeltaE21model}\\[4mm]
\Delta E_{31}&\equiv& E_3 -E_1=
\half \left(\Delta\mu + \Delta b_0 +\sqrt{\Delta\mu^2 + \Delta b_0^2} \right)\,.
\label{DeltaE31model}
\eeqa
\end{subequations}
The most important result for us is that
$|\Delta E_{21}/\Delta E_{31}| \lesssim 10 \%\,$
for $\Delta b_0 \lesssim 0.2\,\Delta\mu$.
In the following discussion, we assume
$0< \Delta b_0 \ll \Delta\mu$, which allows us to use simplified expressions
for the oscillation probabilities; cf. Refs.~\cite{Barger-etal,McKeownVogel}.
(Note that the approximation $\Delta b_0 \gg \Delta\mu >0$ would be equally simple.)

In the context of three-flavor neutrino oscillations,
case 1 of Eq.~(\ref{case123}) with $\theta=0$  may be
relevant to the $\theta_{13}$ mixing angle, which is known to be rather small
from  the CHOOZ \cite{CHOOZ} and
Palo Verde \cite{PaloVerde2001} experiments.
For three-flavor parameters (\ref{standardmassangles})--(\ref{phases}),
large enough beam energy $E_\nu$, and not too large travel distance
($L\Delta b_0 \ll L\Delta\mu \lesssim 1$),
the $\nu_\mu\rightarrow \nu_e$ appearance probability is essentially given by
Eq.~(\ref{PandTheta-case1}) and reads
\begin{subequations}
\label{PandTheta-case13}
\beqa
P_{\mu e}  &\equiv&  P(\nu_\mu\rightarrow \nu_e) \sim
\half\,\sin^2 \left( 2\,\Theta_{13} \right) \;
\sin^2 \left( \, \Delta E_{31}\,L/2\, \right) +
\mathrm{O}\left(  \left(\Delta b_0 \,L \right)^2 \right) \,,
\label{P-case13}\\[4mm]
 2\,\Theta_{13}   &\sim&
\arctan \left( \frac{\Delta b_0}{\Delta\mu} \right)
+ \mathrm{O}\left(  \left(\Delta b_0/\Delta\mu \right)^2 \right) \,,
\label{Theta-case13}
\eeqa
\end{subequations}
with $\Delta E_{31}$ given by expression (\ref{DeltaE31model}).
Analytically, the omitted terms in Eq.~(\ref{P-case13}) would be
negligible if the ``bare'' mixing angle $\theta_{13}$ were small but finite.
But, for the simple model considered with $\theta_{13}=0$ exactly,
the omitted terms  could, in principle,
be important and only a comparison with the full numerical result
can be conclusive (see Sec.~\ref{sec:phenoT2KNOvA}).

Case 2 of Eq.~(\ref{case123}) with $\theta=\pi/4$, on the other hand,
may be relevant to the $\theta_{32}$ mixing angle,
which is known to be nearly maximal from the SK results
\cite{SuperK1998,SuperK2005}.
For three-flavor parameters (\ref{standardmassangles})--(\ref{phases}),
large enough beam energy $E_\nu$, and not too large travel distance $L$,
the $\mu$--type survival probability is essentially  given by
Eq.~(\ref{P-case2}) and reads
\beqa
 P_{\mu \mu} &\equiv& P(\nu_\mu\rightarrow \nu_\mu) \sim        1-
\sin^2 \left( \, \Delta E_{31}\,L/2\,\right) +
\mathrm{O}\left( \Delta b_0\,\Delta\mu\, L^2  \right) \,,
\label{P-case32}
\eeqa
again with $\Delta E_{31}$ given by expression (\ref{DeltaE31model}).
Typically, one expects $\mathrm{O}\left(\Delta b_0/\Delta\mu \right)$
corrections to the probabilities.
There could also  be interesting effects
for an extended version of the model with nonzero phases $\alpha$ and $\beta$,
as mentioned a few lines below Eq.~(\ref{P-case2}).
But, these issues will not be pursued further here.

\section{Phenomenology}
\label{sec:phenomenology}

\subsection{Preliminaries}
\label{sec:preliminaries}

In order to connect with future experiments which aim to determine
or constrain the standard energy-independent mixing angle $\theta_{13}$,
consider the appearance  probability (\ref{PandTheta-case13})
from the simple model (\ref{PXYdef})--(\ref{phases}).
For relatively weak Fermi-point splitting,
$0<\Delta b_0\equiv \Delta b_0^{(31)} \ll \Delta\mu \equiv \Delta m^2_{31}/(2p)$,
the effective mixing angle $\Theta_{13}$ then
has a linear dependence on the neutrino beam energy
$E_\nu \sim p \equiv |\vec p\,|$ and is given by
\beq
\Theta_{13} \sim \frac{E_\nu}{\Delta m^2_{31}}\;\Delta b_0^{(31)}
+ \mathrm{O}\left(  \left(\Delta b_0/\Delta\mu \right)^2 \right)\,,
\label{Theta13linear}
\eeq
since the argument of the arctangent function
in Eq.~(\ref{Theta-case13}) is positive.
If experiment can now establish an upper bound $\Delta\Theta_{13}$
on the variation of $\Theta_{13}(E)$
over an energy range $\Delta E_{\nu\,\mathrm{range}}$,
one obtains an upper bound on $\Delta b_0$.
From Eq.~(\ref{Theta13linear}), one has
\beqa
|\Delta b_0^{(31)}|
&\lesssim&
|\Delta m^2_{31}\;\Delta\Theta_{13}/\Delta E_{\nu\,\mathrm{range}}|\no\\[4mm]
&\approx&
6\times 10^{-13}\;{\rm eV}\,
\left(\frac{|\Delta m^2_{31}|}{2.5\times 10^{-3}\;{\rm eV}^2}\right)\,
\left(\frac{\sin^2 (2\Delta\Theta_{13})}{0.05}\right)^{\! 1/2}\,
\left(\frac{0.5\;{\rm GeV}}{|\Delta E_{\nu\,\mathrm{range}}|}\right)\,,
\label{b0exp}
\eeqa
with more or less realistic values inserted for the planned T2K experiment
\cite{T2K} (for the proposed NO$\nu$A experiment \cite{NOvA},
the energy range could be four times larger, giving a four times
smaller upper bound for the Fermi-point splitting).

A next-generation superbeam \cite{BNL2002,Kobayashi} or
neutrino factory \cite{Geer1997,Blondel2004} could
perhaps reach a $10^2$ times better sensitivity than shown in Eq.~(\ref{b0exp}),
assuming $\sin^2 (2\Delta\Theta_{13})\approx 0.002$
and $\Delta E_{\nu\,\mathrm{range}} \approx 10 \;{\rm GeV}$.
New reactor experiments \cite{Whitepaper}, however, would operate at lower
energies and would have less sensitivity by a factor of $10^2$,
assuming $\sin^2 (2\Delta\Theta_{13}) \approx 0.05$
and $\Delta E_{\nu\,\mathrm{range}} \approx 5 \;{\rm MeV}$.
As mentioned before, the neutrino energy $E_\nu$ must still be large
enough for Eqs.~(\ref{dispersionrelation}),(\ref{2flavormatrix}),
and (\ref{3flavormatrix}) to apply,
which we take to mean $E_\nu \gg 1\;\mathrm{keV}$, based on the
absolute bound $|b_0^{(e)}| \leq 1\;\mathrm{keV}$
reported in Ref.~\cite{DiGrezia-etal},
the conservative differential bounds
$|\Delta b_0^{(ij)}| \leq  10^{-11}\;\mathrm{eV}$ from
Refs.~\cite{KlinkhamerJETPL,Klinkhamer0407200}, and
the absolute bound $\sum_{i} m_{i} \leq 10^{2}\;\mathrm{eV}$
from cosmology \cite{Barger-etal,McKeownVogel}.

Since there could be Fermi-point-splitting effects hiding in the existing neutrino
oscillation data with $|\Delta b_0|$ of the  order of $10^{-12}\;{\rm eV}$
\cite{KlinkhamerJETPL,Klinkhamer0407200},
superbeam experiments \cite{T2K,NOvA,BNL2002,Kobayashi} look the most promising
in the near future, as neutrino  factories may very well
remain in the R\&D phase for at least 10 years.
In Sec.~\ref{sec:phenoT2KNOvA}, we, therefore, expand on
the  appearance  probability $P(\nu_\mu\rightarrow \nu_e)$
at forthcoming off-axis superbeam experiments which can be expected
to have relatively good control of the backgrounds.
In Sec.~\ref{sec:phenoMINOS}, we also discuss this appearance
probability for the on-axis MINOS experiment which can have relatively
high neutrino energies.

\subsection{Model results near the first oscillation maximum}
\label{sec:phenoT2KNOvA}

In this subsection, we consider the appearance  probability
$P_{\mu e} \equiv P(\nu_\mu\rightarrow \nu_e)$ from
the simple model (\ref{PXYdef})--(\ref{phases}) with
both mass-square differences and timelike Fermi-point splittings.
Experimentally, there are two important parameters to determine
from the measured values of the quantity $P_{\mu e}=P_{\mu e}(L,E_\nu)$,
namely $\Delta m^2_{31}$ and $\Delta b_0^{(31)}$, where $|\Delta b_0^{(31)}|$
is assumed to be less than $|\Delta m^2_{31}|/(2E_\nu)$ for typical energies
$E_\nu$. Perhaps the simplest way to determine or constrain $|\Delta b_0^{(31)}|$
would be to place the far detector close to the first oscillation maximum
corresponding to the average energy $\widebar{E}_\nu$.
The near detector close to the source determines the
initial  $\nu_\mu$ flux. The idea is
then to measure the $\nu_e$ energy spectrum in the far detector
and to compare with the standard predictions.

The shape of $P_{\mu e} $ vs. $E_\nu$  for our combined
mass-difference and Fermi-point-splitting
(MD+FPS) model is, indeed, different
from the one of the standard mass-difference (MD) model. Defining
\mathindent=0.25cm
\begin{subequations}\label{def}
\beqa
\widebar{L} &\equiv& \pi\;\frac{2\,\widebar{E}_\nu}{|\Delta m^2_{31}|}
\approx 295\;\mathrm{km}\,
\left( \frac{\widebar{E}_\nu}{0.5948\;\mathrm{GeV}} \right)\,
\left( \frac{2.5\times 10^{-3}\;{\rm eV}^2}{|\Delta m^2_{31}|} \right),
\label{defLbar}
\\[4mm]
\widebar{\Theta}_{13} &\equiv&
\widebar{E}_\nu\,\frac{|\Delta b_0^{(31)}|}{|\Delta m^2_{31}|}
\approx 0.0952 \,
\left( \frac{\widebar{E}_\nu}{0.5948\;\mathrm{GeV}} \right)
\left( \frac{|\Delta b_0^{(31)}|}{4\times 10^{-13}\;\mathrm{eV}} \right)
\left( \frac{2.5\times 10^{-3}\;{\rm eV}^2}{|\Delta m^2_{31}|} \right),
\label{defTheta13bar}
\\[4mm]
l &\equiv& L/\widebar{L} \,,
\label{defl}
\\[4mm]
x &\equiv& E_\nu/\widebar{E}_\nu \,,
\label{defx}
\eeqa
\end{subequations}
\mathindent=\mathindentnum
the approximate model probability (\ref{PandTheta-case13})  becomes
\begin{subequations}\label{PandTheta13model}
\beqa
P_{\mu e}^\mathrm{\,MD+FPS}(l,x) &\sim&
\half\,\sin^2 ( 2\,\Theta_{13} )\;
\sin^2 \left(
\frac{\pi}{4}
\left[\: x^{-1} + 2\,\widebar{\Theta}_{13} +
\sqrt{ x^{-2} + \left(2\,\widebar{\Theta}_{13}\right)^2} \;\right] \, l
\, \right) ,
\label{Pmodel} \\[4mm]
 2\,\Theta_{13}
&\sim& \arctan (2\,\widebar{\Theta}_{13}\,x)  \,.
\label{Theta13model}
\eeqa
\end{subequations}
For comparison, the standard mass-difference result for
$m_1^2=m_2^2 \ne m_3^2$  and $\theta_{32}=\pi/4$
is given by \cite{GribovPontecorvo,
BilenkyPontecorvo1976,BilenkyPontecorvo1978,Barger-etal,McKeownVogel}
\beqa
P_{\mu e}^\mathrm{\,MD}(l,x) &=&
\half\, \sin^2 (2\theta_{13})\;
\sin^2 \left(\,\frac{\pi}{2\,x}\; l\, \right)\,,
\label{Pstandard}
\eeqa
with constant mixing angle $\theta_{13}$.

\begin{figure*}[t]
\begin{center}
\includegraphics[width=8.5cm]{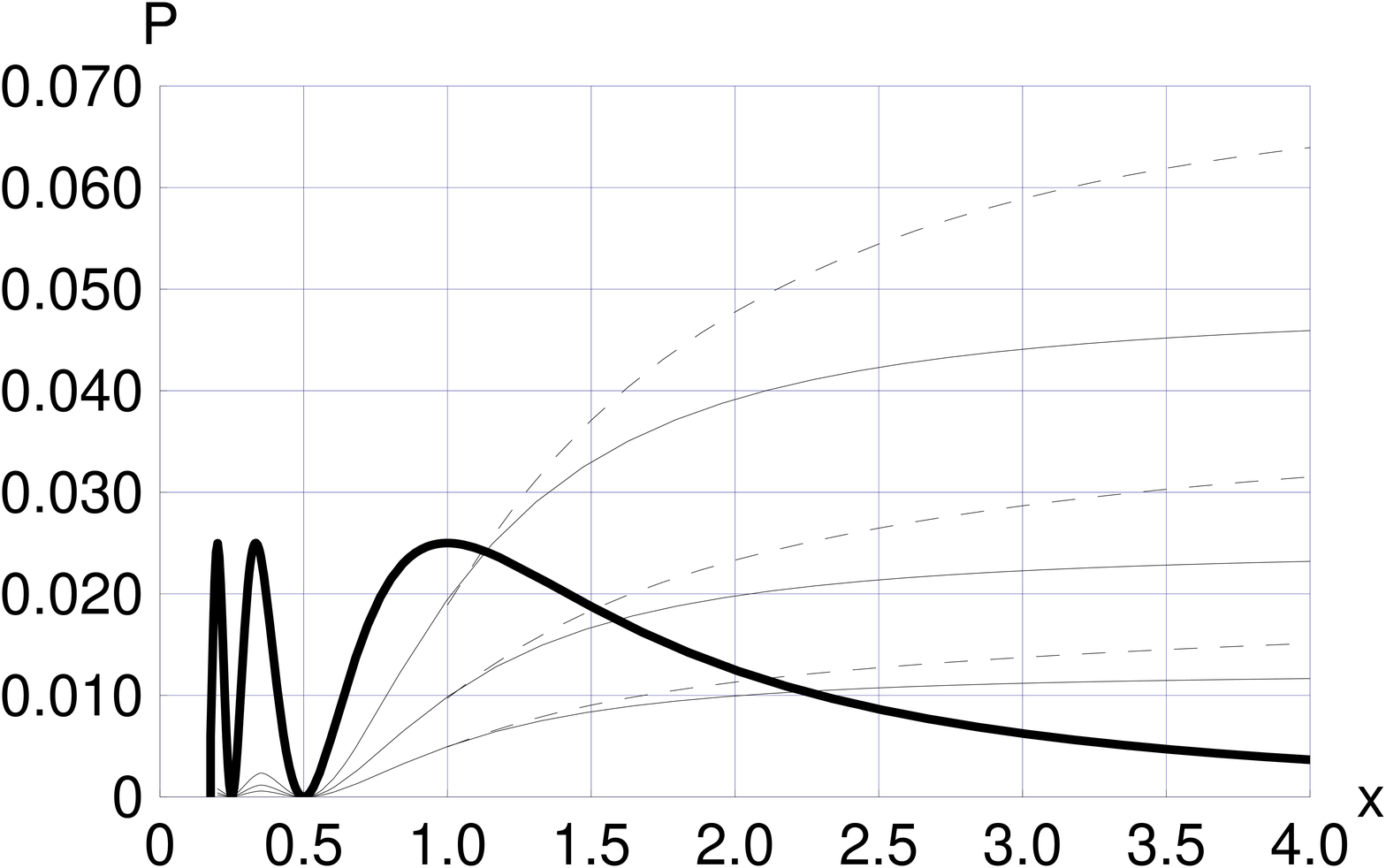}
\end{center}
\vspace*{-0.5cm}
\caption{Fixed-distance probabilities
$P(\nu_\mu\rightarrow \nu_e)$  at $l \equiv L/\widebar{L}=1$
vs. dimensionless energy $x\equiv E_\nu/\widebar{E}_\nu$
for the simple model (\ref{PXYdef})--(\ref{phases})
with both mass differences
and Fermi-point splittings [thin broken and solid curves]
and for the standard mass-difference model without Fermi-point splittings
[heavy solid curve with maxima $P=0.025$ at $x \leq 1$,
the plot  being suppressed for $x \leq 0.175\,$].
The thin broken curves and the  heavy solid curve are given by the analytic
expressions (\ref{PandTheta13model}) and  (\ref{Pstandard}), respectively,
with parameters $\sin^2 (2\widebar{\Theta}_{13}) =0.01,\, 0.02, \,0.04$,
and $\sin^2 (2\theta_{13})=0.05$. The thin broken curves
are only approximate for $x \widebar{\Theta}_{13}=\mathrm{O}(1)$
and the corresponding thin solid curves give the full numerical results.}
\label{fig1T2K}
\end{figure*}

Figure \ref{fig1T2K} compares the shape
of the model probability (\ref{PandTheta13model})  to the standard
result (\ref{Pstandard}), both evaluated  at $l\equiv L/\widebar{L}=1$.
The approximate model probabilities  from Eq.~(\ref{PandTheta13model})
[thin broken curves in Fig.~\ref{fig1T2K}] are quite reliable for $x\approx 1$ but
overshoot by some $40\, \%$ for larger values of $x$,
as follows by comparison to the full numerical results
[thin solid curves  in Fig.~\ref{fig1T2K}] which monotonically approach
a constant value as $x \to \infty$. Note, however, that this asymptotic
value $P=\half\,\sin^2 ( \pi\,\widebar{\Theta}_{13} )=
\half\,\sin^2 (\Delta b_0\, \widebar{L}/2 )$ is correctly
reproduced by the approximation (\ref{Pmodel}) for $l=1$ and $1/x=0$.
The analytic expression (\ref{Pmodel}) can, therefore, be used
to get rough estimates.
(For corresponding probabilities in another model,
  see Fig.~8 in Ref.~\cite{Klinkhamer0407200}.)   

\begin{figure*}[t]
\begin{center}
\includegraphics[width=8.5cm]{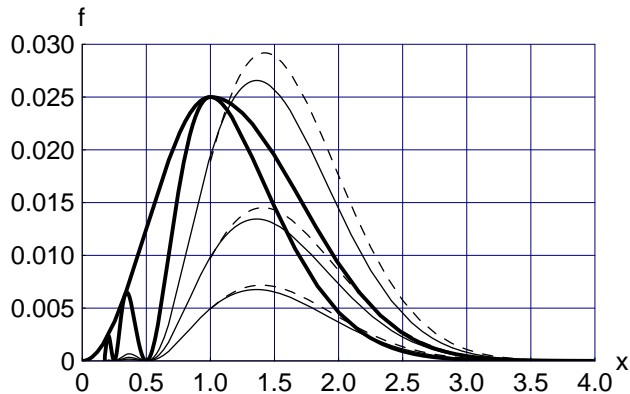}
\end{center}
\vspace*{-0.5cm}
\caption{Electron--type neutrino energy spectra $f_e(x)$
at $l \equiv L/\widebar{L}=1$
from an initial muon--type spectrum (\ref{initialspectrum}) at $l=0$
and probabilities $P(\nu_\mu\rightarrow \nu_e)$ of Fig.~\ref{fig1T2K}.
The initial $\nu_\mu$ spectrum multiplied by a constant factor $0.025$
is shown as the upper  heavy solid curve.
The $\nu_e$ spectra at $l=1$ are shown as
the lower heavy solid curve for the standard mass-difference model
($\sin^2 \,2\theta_{13}=0.05$) and
as the thin broken and solid curves for the simple model with both
mass differences ($\theta_{13}=0$) and Fermi-point splittings
(see Fig.~\ref{fig1T2K} for further details).}
\label{fig2T2K}
\end{figure*}

The  $\nu_e$ energy spectrum at $L=\widebar{L}$ depends on the initial
$\nu_\mu$ spectrum at $L= 0$, assuming negligible contamination by $\nu_e$'s.
(Note that, for a real experiment, backgrounds are better controlled in the
off-axis configuration than in the on-axis configuration \cite{T2K,NOvA}.)
As an example, we take for this initial $\nu_\mu$ spectrum the
following function [cf. upper  heavy solid curve  in Fig.~\ref{fig2T2K}\,]:
\beq
\left.  f_\mu(x)\right|_{\,L= 0}=
\left\{ \begin{array}{ll}
\sin^2(x\,\pi/2)\,, & \;\;\mbox{for $x \in [0,1)$}\,,\\[4mm]
\exp\left[-(x-1)^2 \,\right]\,, & \;\;\mbox{for $x \in [1,\infty)$}\,,
\end{array} \right.
\label{initialspectrum}
\eeq
in terms of the dimensionless energy $x$ from Eq.~(\ref{defx}) and
with an arbitrary normalization.
The $\nu_e$ spectrum at $L=\widebar{L}$  (or $l=1$) is then given by
\beq
  \left.  f_e(x)\right|_{\,L=\widebar{L}}=
\left.  f_\mu(x)\right|_{\,L= 0} \;
P_{\mu e}(1,x) \,.
\label{finalspectrum}
\eeq
Figure \ref{fig2T2K} shows the resulting $\nu_e$ energy spectra:
the lower heavy solid curve corresponds to the probability
(\ref{Pstandard}) of the standard mass-difference model
and the thin solid (broken) curves correspond to the numerical (approximate)
probabilities of the simple model with both
mass differences and Fermi-point splittings, the
approximate probability being given by Eq.~(\ref{Pmodel}).

The reader is invited to compare the solid curves of
Fig.~\ref{fig2T2K} with Figs. 6b and 13a  of Ref.~\cite{T2K} for T2K,
which give the initial $\nu_\mu$ spectrum and
the expected $\nu_e$ signal for mass-difference oscillations with
$\sin^2 (2\theta_{13})=0.1$ (i.e., twice the value used in our figures).
T2K would appear to be able to measure the
oscillation probability (\ref{PandTheta13model}) for
$\sin^2(2\widebar{\Theta}_{13})= 0.04$, which
corresponds to $|\Delta b_0^{(31)}| \approx
4\times 10^{-13}\;\mathrm{eV}$, according to
Eq.~(\ref{defTheta13bar}) for $\widebar{E}_\nu= 0.6\;\mathrm{GeV}$
and $\Delta m^2_{31}=2.5\times 10^{-3}\;{\rm eV}^2$.
The proposed NO$\nu$A experiment \cite{NOvA}
would be sensitive to  approximately  four times smaller values of
the Fermi-point splitting, $|\Delta b_0^{(31)}| \approx 10^{-13}\;\mathrm{eV}$,
because of the approximately  four times higher energy,
even though $L \approx 810\;\mathrm{km}$ is a bit short of
$\widebar{L} \approx 1200 \;\mathrm{km}$.
(For a generalized model with $\theta_{13}=0$ exactly and
$\chi_{13}$ arbitrary, similar sensitivities hold for the quantity
$|\Delta b_0^{(31)}\,\sin 2\chi_{13}|\,$.)

The sensitivity is one issue but the ability to choose between
different models is another. In fact,
it would not be easy to distinguish between the standard
mass-difference model with $\sin^2 2\theta_{13}=0.05$ [lower
heavy solid curve of Fig.~\ref{fig2T2K}\,] and
the combined mass-difference and Fermi-point-splitting model with
$\theta_{13}=0$ and $\sin^2 (2\widebar{\Theta}_{13})=0.04$
[upper thin solid curve of Fig.~\ref{fig2T2K}\,].
Possible signatures of the simple model  (\ref{PXYdef})--(\ref{phases})
for the $\nu_e$ energy spectrum at $L=\widebar{L}$
would be a shift of the initial $x=1$ peak to $x \approx 1.4$,
practically zero activity for $x \lesssim 0.5$, and
an enhanced signal for $x \gtrsim 2$.

\begin{figure*}
\begin{center}
\includegraphics[width=8.5cm]{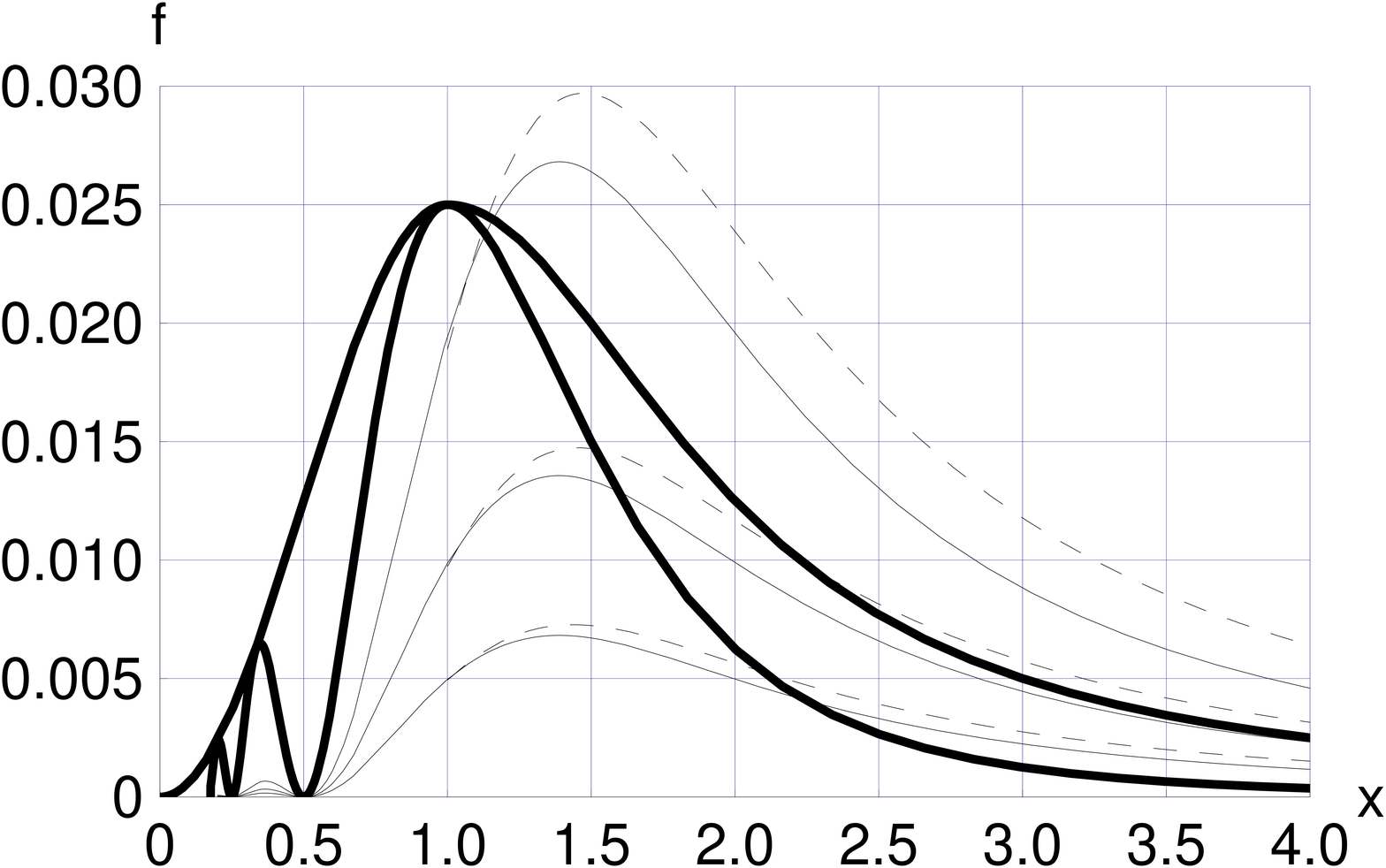}
\end{center}
\vspace*{-0.5cm}
\caption{Same as Fig.~\ref{fig2T2K}, but now for initial muon-type
neutrino energy spectrum (\ref{initialspectrumtail}).}
\label{fig3T2K}
\end{figure*}

Expanding on this last point, it is clear that an initial energy spectrum with a
pronounced high-energy tail would be advantageous. As an example, we take
the following function [cf. upper  heavy solid curve  in Fig.~\ref{fig3T2K}\,]:
\beq
\left.  \widetilde{f}_\mu(x)\right|_{\,L= 0}=
\left\{ \begin{array}{ll}
\sin^2(x\,\pi/2)\,, & \;\;\mbox{for $x \in [0,1)$}\,,\\[4mm]
1/ \left[1+    (x-1)^2 \,\right] \,, & \;\;\mbox{for $x \in [1,\infty)$}\,,
\end{array} \right.
\label{initialspectrumtail}
\eeq
which has an area larger than the one from Eq.~(\ref{initialspectrum})
by a factor $(1+\pi)/(1+\sqrt{\pi})\approx 1.5$.
Figure \ref{fig3T2K} shows
the resulting $\nu_e$ energy spectra from Eq.~(\ref{finalspectrum}) with
$f_\mu(x)$ replaced by $\widetilde{f}_\mu(x)$. In this case,
the new-physics signal at $x \approx 2$ would be quite strong, being
approximately three times larger for
the combined mass-difference and Fermi-point-splitting model with
$\theta_{13}=0$ and  $\sin^2 (2\widebar{\Theta}_{13}) =0.04$
[upper thin solid curve of Fig.~\ref{fig3T2K}\,] than for  the standard
mass-difference model with $\sin^2 (2\theta_{13})=0.05$ [lower
heavy solid curve of Fig.~\ref{fig3T2K}\,].

In order to distinguish Fermi-point-splitting
effects from mass-difference effects, it is clearly preferable
to use a  (pure) $\mu$--type neutrino beam with a broad energy
spectrum and high average energy. For this reason, we take a closer
look at the potential of the current MINOS experiment in the next subsection.

\subsection{Model results short of the first oscillation maximum}
\label{sec:phenoMINOS}

In this subsection, we give model results which may be relevant
to the on-axis MINOS experiment \cite{MINOS1998,MINOS2005}
with baseline $L=735\;\mathrm{km}$. In order to be specific, we
focus on the medium energy (ME) mode of MINOS with an average beam
energy of approximately $\widebar{E}_\nu= 7.5\;\mathrm{GeV}$,
but the results are qualitatively the same for
the  low-energy (LE) and  high-energy (HE) modes
with $\widebar{E}_\nu$ approximately equal to $3.75\;\mathrm{GeV}$ and
$15\;\mathrm{GeV}$, respectively.
The average energy $\widebar{E}_\nu= 7.5\;\mathrm{GeV}$ corresponds to
a length scale
$\widebar{L}\approx 3720 \;\mathrm{km}$, according to Eq.~(\ref{defLbar})
for $\Delta m^2_{31}=2.5\times 10^{-3}\;{\rm eV}^2$.

\begin{figure*}[t]
\begin{center}
\includegraphics[width=8.5cm]{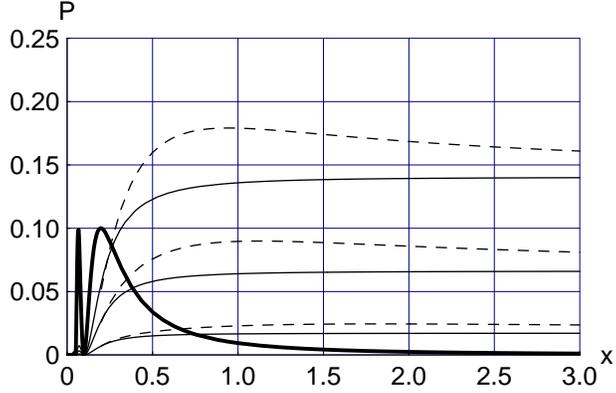}
\end{center}
\vspace*{-0.5cm}
\caption{Fixed-distance probabilities
$P(\nu_\mu\rightarrow \nu_e)$ at $l \equiv L/\widebar{L}=735/3720$ vs.
dimensionless energy $x\equiv E_\nu/\widebar{E}_\nu$
for the simple model (\ref{PXYdef})--(\ref{phases})
with both mass differences and Fermi-point splittings.
The thin broken curves are given by the analytic
expression (\ref{PandTheta13model})
with parameters $\widebar{\Theta}_{13}=0.3,\, 0.6, \,0.9$.
The thin broken curves are only approximate
for $x \widebar{\Theta}_{13}=\mathrm{O}(1)$
and the corresponding thin solid curves give the full numerical results.
The heavy solid curve gives, for comparison, the standard mass-difference
probability (\ref{Pstandard}) for $\sin^2 (2\theta_{13})=0.2$ and
$l=735/3720$.}
\label{fig4MINOS}
\end{figure*}

The approximate probability for the simple model (\ref{PXYdef})--(\ref{phases})
is again given by Eq.~(\ref{PandTheta13model}).
Figure \ref{fig4MINOS} shows
the approximate  probabilities and the full numerical results
as thin broken and thin solid curves, respectively, for model
parameters $l \equiv L/\widebar{L}=735/3720$ and
$\widebar{\Theta}_{13}=(0.3,\, 0.6, \,0.9)$, which
correspond to $\Delta b_0 \approx (1,\,2,\,3) \times 10^{-13}\;\mathrm{eV}$,
according to Eq.~(\ref{defTheta13bar})
for $\widebar{E}_\nu= 7.5\;\mathrm{GeV}$
and $\Delta m^2_{31}=2.5\times 10^{-3}\;{\rm eV}^2$.
The results for the  LE and HE modes of MINOS
are similar for $x \gtrsim 1$, because the asymptotic value
$P=\half\,\sin^2 (\Delta b_0\, L/2 )$ is independent of the beam energy
(for corresponding probabilities in another model,
  see Fig.~7 in Ref.~\cite{Klinkhamer0407200}).       

Note that the standard mass-difference probability (\ref{Pstandard})
at $x=1$ is less than $1\,\%$ for $l=735/3720$
and $\sin^2 (2\theta_{13})=0.2$, which is just above the $90\,\%$
CL limit (analysis A at $\Delta m^2=2.5\times 10^{-3}\;{\rm eV}^2$)
from CHOOZ  \cite{CHOOZ}.
This standard probability is shown as the heavy solid curve
in Fig.~\ref{fig4MINOS}.

\begin{figure*}
\begin{center}
\includegraphics[width=8.5cm]{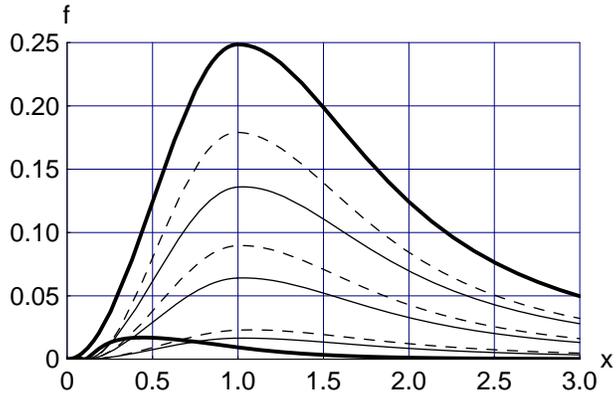}
\end{center}
\vspace*{-0.5cm}
\caption{Electron--type neutrino energy spectra $f_e(x)$ at
$l \equiv L/\widebar{L}=735/3720$
from an initial muon--type spectrum (\ref{initialspectrumtail})
at $l=0$ and probabilities $P(\nu_\mu\rightarrow \nu_e)$
of Fig.~\ref{fig4MINOS}.
The initial $\nu_\mu$ spectrum multiplied by a constant factor $0.25$
is shown as the upper heavy solid curve. The $\nu_e$ spectra are shown
as the thin solid curves [approximate values as thin broken curves]
for the simple model with both
mass differences ($\theta_{13}=0$) and Fermi-point splittings
(see Fig.~\ref{fig4MINOS} for further details).
The lower heavy solid curve gives, for comparison,
the $\nu_e$ spectrum from the standard mass-difference
mechanism with $\sin^2 (2\theta_{13})=0.2$ and $l=735/3720$.}
\label{fig5MINOS}
\end{figure*}

Figure \ref{fig5MINOS} gives
the resulting $\nu_e$ energy spectra for an initial $\mu$--type
spectrum (\ref{initialspectrumtail}). Again, the results are
qualitatively the same for the LE and HE energy-modes of MINOS,
but the expected event rates for the HE mode would be larger than
for the ME mode and the spectrum would also be somewhat broader.

To summarize,
if MINOS in the ME mode is able to place an upper limit
on the appearance probability $P(\nu_\mu\rightarrow \nu_e)$
of the order of $5\,\%$, this would correspond
to an upper limit on $|\Delta b_0^{(31)}\,\sin 2\chi_{13}|$
of the order of $2\times 10^{-13}\;\mathrm{eV}$ (and a factor two better
in the HE mode).
Moreover, any signal at the $5\,\%$ level or more for $x \gtrsim 1$
would indicate
nonstandard (Fermi-point-splitting?) physics,  since, as mentioned above,
the  standard mass-difference probability
can be expected to be at most $1\,\%$.

\section{Conclusion}
\label{sec:conclusion}

In this article, we have considered a simple neutrino-oscillation
model (\ref{PXYdef})--(\ref{phases}) with both nonzero mass-square
differences ($m^2_{1} =m^2_{2} \ne m^2_{3}$) and
timelike Fermi-point splittings ($b_0^{(1)} =b_0^{(2)} \ne b_0^{(3)}$),
together with a combined bi-maximal and trimaximal mixing pattern.
We expect our basic conclusions to carry over to a more general model
with, for example, nonzero phases
$\delta,\epsilon,\alpha,\beta,$ and all values of
$m^2_{n}$ and $b_0^{(n)}$ different. As mentioned in the Introduction,
the main goal of the present article has been  to point out a
possible energy dependence of \emph{all} neutrino-oscillation parameters
(for three flavors: three mixing angles $\Theta_{ij}$ and one phase
$\Delta$, in addition to the two eigenvalue differences $\Delta E_{ij}$).

Taking the $\theta_{13}$ angle of the mass-square matrix in the simple model
to be $0$ (or very small) and the  $\chi_{13}$ angle of
the Fermi-point-splitting matrix to be $\pi/4$ (or very close to maximal),
we have calculated the energy-dependent effective mixing angle $\Theta_{13}$
for the $\nu_\mu \to \nu_e$ appearance probability.
This probability $P(\nu_\mu \to \nu_e)$ is approximately
given by Eq.~(\ref{PandTheta-case13}) and,
near the first oscillation maximum, by Eq.~(\ref{PandTheta13model}) for
$l \equiv L/\widebar{L}=1$ and  Fig.~\ref{fig1T2K}.
The resulting $\nu_e$ energy spectra are shown in Figs.~\ref{fig2T2K}
and \ref{fig3T2K} for different initial $\nu_\mu$ energy spectra.
Similar results at a smaller value of the dimensionless
distance $l$ have been presented in
Figs.~\ref{fig4MINOS}  and \ref{fig5MINOS}.

These results show that, even for
a (simple) model with Fermi-point splittings hiding behind mass differences,
a value $\theta_{13}\approx 0$ would allow for a detection of
$|\Delta b_0^{(31)}\,\sin 2\chi_{13}|$ at the level
of  $10^{-13}\;\mathrm{eV}$ in a NO$\nu$A--type experiment.
But, it is also  possible that MINOS can already
place tight upper bounds on (or make the discovery of)
Fermi-point splitting, provided the backgrounds are under control.

\noindent\section*{ACKNOWLEDGMENTS}

Part of this work was done at the Center for Theoretical Physics of the
Massachusetts Institute of Technology
and discussions with members and visitors are gratefully acknowledged.
The author also thanks Milind
Diwan and Jacob Schneps for further discussions on the MINOS experiment
and NO$\nu$A  proposal.

\end{document}